\documentclass{article}

\usepackage{amsmath}
\usepackage{PRIMEarxiv}

\usepackage[utf8]{inputenc} 
\usepackage[T1]{fontenc}    
\usepackage{hyperref}       
\usepackage{url}            
\usepackage{booktabs}       
\usepackage{amsfonts}       
\usepackage{nicefrac}       
\usepackage{microtype}      
\usepackage{lipsum}
\usepackage{graphicx}
\graphicspath{{media/}}     

\title{Photonic KAN: A Kolmogorov-Arnold Network Inspired Efficient Photonic Neuromorphic Architecture
}

\author{%
    \textbf{Yiwei Peng\textsuperscript{\rm 1, \rm *}},
    \textbf{Sean Hooten \textsuperscript{\rm 1, \rm *}},
    \textbf{Xinling Yu\textsuperscript{\rm 1}},
    \textbf{Thomas Van Vaerenbergh\textsuperscript{\rm 2}},
    \textbf{Yuan Yuan\textsuperscript{\rm 1}},
    \textbf{Xian Xiao\textsuperscript{\rm 1}},\\
   \textbf{Bassem Tossoun\textsuperscript{\rm 1}},
   \textbf{Stanley Cheung\textsuperscript{\rm 1}},
   \textbf{Marco Fiorentino\textsuperscript{\rm 1}},
   \textbf{Raymond Beausoleil\textsuperscript{\rm 1}}\\
  \textsuperscript{\rm 1} Hewlett Packard Labs, Hewlett Packard Enterprise\\
  \textsuperscript{\rm 2} Hewlett Packard Labs, HPE Belgium\\
  \textsuperscript{\rm *} Equal Contributions
}

\begin{document}
\maketitle

\begin{abstract}
    Kolmogorov-Arnold Networks (KAN) models were recently proposed and claimed to provide improved parameter scaling and interpretability compared to conventional multilayer perceptron (MLP) models. Inspired by the KAN architecture, we propose the Photonic KAN -- an integrated all-optical neuromorphic platform leveraging highly parametric optical nonlinear transfer functions along KAN edges. In this work, we implement such nonlinearities in the form of cascaded ring-assisted Mach-Zehnder Interferometer (MZI) devices. This innovative design has the potential to address key limitations of current photonic neural networks. In our test cases, the Photonic KAN showcases enhanced parameter scaling and interpretability compared to existing photonic neural networks. The photonic KAN achieves approximately 65× reduction in energy consumption and area, alongside a 50× reduction in latency compared to previous MZI-based photonic accelerators with similar performance for function fitting task. This breakthrough presents a promising new avenue for expanding the scalability and efficiency of neuromorphic hardware platforms.
\end{abstract}


\section{Introduction}
Multi-Layer Perceptrons (MLPs) are fully-connected feedforward neural networks that consist of multiple layers of nodes, and are considered the backbone of today's deep learning models. Each node processes information by applying a fixed activation function to the weighted sum of its inputs. They can theoretically approximate any continuous function, given enough layers and neurons \cite{rumelhart1986learning,hornik1989multilayer}. This capability allows for widespread use in diverse deep learning tasks, such as classification, regression, and processing of natural language. While versatile, MLPs also have limitations, including challenges in interpreting learned representations and difficulties in scaling the network effectively.

Kolmogorov-Arnold Networks (KANs), introduced in \cite{liu2024kan}, present a compelling alternative to traditional MLPs. A key distinction lies in KANs' ability to learn activation functions on edges, offering greater flexibility. The original KAN model utilizes B-splines to construct activation functions, replacing linear weight parameters with adaptable spline-based functions. The advantages of this approach, including scalability, flexibility, efficiency, and interpretability, have spurred further investigation into their potential \cite{yu2024kan,shukla2024comprehensive,bodner2024convolutional}.

Digital electronics hardware currently dominates the AI accelerator market. Nevertheless, traditional electrical digital computing platforms are encountering substantial obstacles as transistor-based chips struggle to deliver further performance improvements without consuming excessive power.  Furthermore, digital electronic computing is inherently constrained by its clock speed, leading to restricted inference bandwidth and significant latency bottlenecks \cite{chang2010practical,dreslinski2010near}.

Integrated photonic accelerators performing analog processing are emerging as a powerful alternative, promising a next-generation computing platform with ultra-high speed, massive parallelism, energy efficiency, and real-time processing. Researchers are actively exploring various optical systems, including Mach-Zehnder Interferometers (MZI) meshes \cite{shen2017deep,zhou2022photonic, yuan2023low}, microring resonator (MRR) crossbars \cite{ohno2022si}, and coherent MRR networks \cite{wang2023microring}, to implement MLP layers in the photonic domain. These accelerators are primarily optimized for shallow, weight-static neural networks. Despite their potential, current photonic accelerators face significant challenges that limit their practical usefulness:

\begin{enumerate}
    \item Photonic accelerators lack depth due to the absence of efficient on-chip nonlinear activation functions (NAFs). Current artificial neural network (ANN) architectures, relying on cascaded MLPs and NAFs, are challenging to implement directly in the all optical domain. Traditional optical neural networks (ONNs), requiring optical-electrical-optical (OEO) conversions for nonlinear operations, introduce substantial latency, high power consumption (0.1 W per channel \cite{williamson2019reprogrammable,bandyopadhyay2022single}), and bulky footprint. This problem is exacerbated in deep neural networks, where multiple OEO conversions diminish the advantages of photonic accelerators.
    \item Photonic accelerators also have restricted network width, which makes it struggle with extensibility due to inherent design limitations. For instance, MZI-based ONN typically relies on Singular Value Decomposition (SVD) techniques combining meshes of cascaded 2×2 MZIs. An $N$×$N$ matrix necessitates $N^2$ MZI nodes with a minimum and maximum optical path of ($N$ + 1) and (2$N$ + 1) MZI, respectively, resulting in significant accumulated loss as $N$ increases \cite{giamougiannis2023coherent}. This necessitates higher laser power or additional amplifiers to compensate, increasing energy consumption. Current ANN architectures pose challenges for scaling photonic accelerators to tackle more complex problems.
\end{enumerate}

Existing photonic accelerators struggle with scalability and implementing traditional ANN architectures. For example, while a single NVIDIA H100 GPU can handle a popular language model GPT-2 XL with 1.5 billion parameters \cite{emani2024toward}, current state-of-the-art optical neural networks are limited to a scale of 64 x 64 or smaller \cite{ramey2020silicon}. To overcome these challenges, we introduce Photonic KAN, the first customized photonic accelerator designed for Kolmogorov–Arnold Networks.

\section{Preliminaries}
\label{prelim}

The power of KANs lies in their distinctive architecture. In contrast to traditional MLPs, which rely on fixed activation functions at nodes, KANs employ adaptable activation functions on network edges. Leveraging the principles of the Kolmogorov-Arnold Representation Theorem \cite{KAN_RT}, the learning of a complex, high-dimensional function can be simplified to the learning of a manageable number of one-dimensional functions. This approach empowers KANs with a highly flexible and adaptable architecture, capable of dynamically adjusting to intricate data patterns. As a result, a KAN layer with $n_{in}$-dimensional inputs and $n_{out}$-dimensional outputs can be defined as a matrix of 1D functions:
\begin{equation}
\Phi = \{\phi_{q,p}\}, \quad p = 1,  \ldots, n_{in}, \quad q = 1, \ldots, n_{out}
\end{equation}
where the functions $\phi_{q,p}$ are univariate functions having trainable parameters. Particularly, each function can be defined as a spline, Chebyshev orthogonal polynomial, or radial basis function. We denote the $j$-th activation function of $i$-th neuron in the $l$-th layer as $\phi_{l,i,j}$. The activation value of the ($l + 1$, $j$) neuron is simply the sum of all incoming activation functions $x_{l+1,j} = \sum_{i=1}^{n_l} \phi_{l,j,i}(x_{l,i})$. Each layer’s transformation, $\phi_{l}$, acts on the input $x_{l}$ to produce the next layer’s input $x_{l+1}$ in matrix form is described as \cite{liu2024kan}:
\begin{equation}
     x_{l+1} = \Phi_l(x_l) = \left( \begin{array}{ccc}
\phi_{l,1,1}(\cdot) & \cdots & \phi_{l,1,n_l}(\cdot) \\
\vdots & \ddots & \vdots \\ 
\phi_{l,n_{l+1},1}(\cdot) & \cdots & \phi_{l,n_{l+1},n_l}(\cdot)
\end{array} \right) x_l.
\label{1}
\end{equation}
A general KAN network consists of $L$ layers, and hence may be expressed as:
\begin{equation}
\text{KAN}(x) = (\Phi_{L-1} \circ \Phi_{L-2} \circ \cdots \circ \Phi_0)(x),
\end{equation}
which is analogous to increasing the depth of MLPs.

\section{Proposed Photonic KAN Design}
\label{KAN_Design}
\begin{figure}[h!]
    \centering
    \includegraphics[scale=0.28]{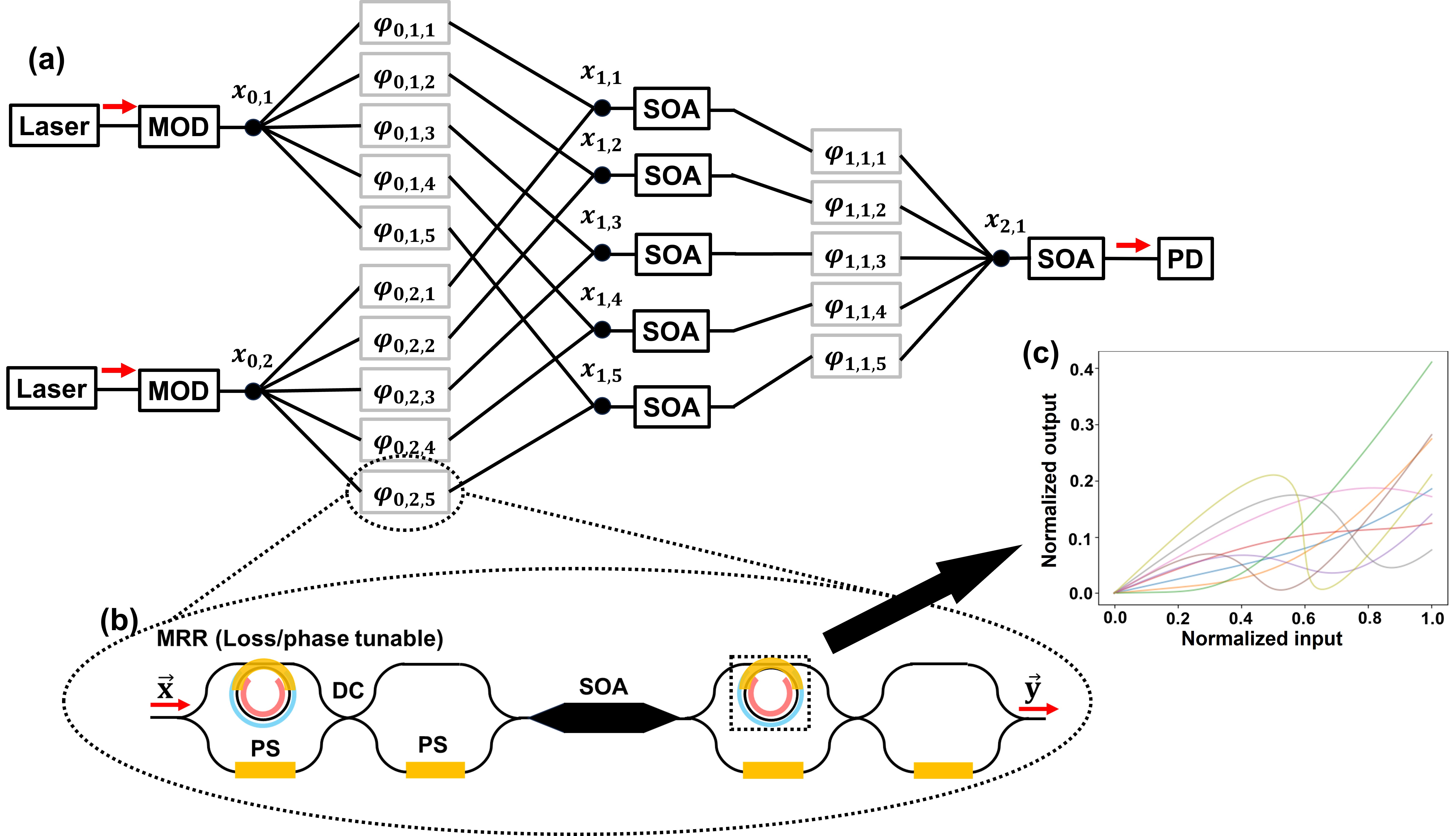}
    \caption{(a) Our proposed photonic KAN. (b) Enlarged D-RAMZI unit cell. (c) Distinct nonlinear functions of single MRR.}
    \label{KAN_arch}
\end{figure}

Here we propose a photonic neuromorphic architecture that leverages tunable nonlinear functions along network edges, inspired by the KAN discussed previously. As shown in Fig. \ref{KAN_arch}(a), the laser light first undergoes input vector modulation, then passes through a 1:N splitter, which can be built using 1:2 splitters. If the laser power is insufficient, additional lasers can be cascaded to boost it. At the core of our photonic KAN, we will employ a single-input, single-output all-optical nonlinear unit. Our design introduces a learnable basic block -- in this case the MRR-assisted MZI (RAMZI) unit -- capable of generating diverse nonlinear functions through parameter adjustments \cite{peng2023all}. We will experimentally demonstrate this unit's functionality and create a behavioral model based on temporal coupled-mode theory (CMT). By leveraging RAMZI units, the network replaces standard linear edge weights with edges that represent adaptable and learnable functions.

Nonlinearity originates from the high-Q MRR within the RAMZI structure. Free-carrier dispersion (FCD) accumulates within the MRR, inducing a nonlinear phase shift that the MZI converts into a nonlinear transmission response. The structure's dynamic behavior is modeled using rate equations and coupled-mode theory \cite{peng2023all,jha2020reconfigurable}. The dynamic equations can be simplified with light amplitude in the MRR $a$ and a free carrier density $N_c$ as:

\begin{equation}\label{eq4}
\frac{da}{dt} = -(j\Delta \omega + \gamma_L)a - j \eta_{fc} N_c a - j \mu \sqrt{P_{in}}
\end{equation}
\begin{equation}\label{eq5}
\frac{dN_c}{dt} + \frac{N_c}{\tau_{fc}} = \xi |a|^4
\end{equation}

where $\Delta \omega$ is the detuning between input signal frequency and resonance frequency; $\gamma_L$, $\mu$, $P_{\text{in}}$ and $\tau_{\text{fc}}$ represents linear loss, field coupling coefficient, input power and free carrier lifetime; $\eta_{\text{fc}}$ and $\xi$ relate to free carrier effect and two-photon absorption, respectively \cite{van2016optical}. Due to its negligible impact compared to the free carrier effect, the Kerr effect in silicon is omitted from our equations.

\begin{figure}[h!]
    \centering
    \includegraphics[scale=0.44]{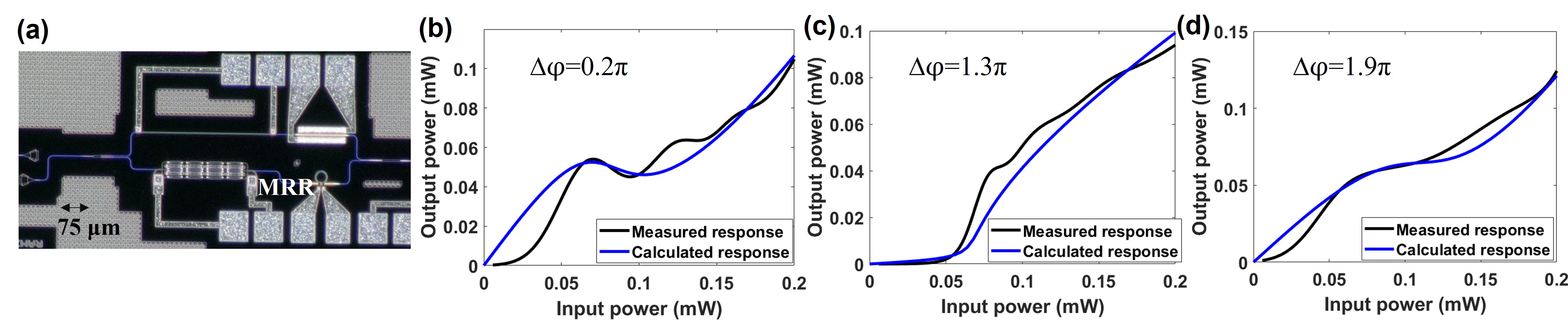}
    \caption{(a) Micrograph of the fabricated RAMZI. (b)-(d) Measured and simulated  NAFs for RAMZI at various $\Delta \phi$.}
    \label{Measure}
\end{figure}

When operating at low input power, the MRR's output is minimal near its initial resonant wavelength. However, as input power rises, the resonance undergoes a blueshift, resulting in a brief decrease in output followed by a rapid increase. Crucially, the specific nonlinear function of the MRR can be tailored by adjusting loss and detuning resonance. These parameters can be individually tuned via carrier injection and thermal effects. Metal–oxide–semiconductor capacitor (MOSCAP) phase tuning presents a promising alternative, eliminating static power consumption and boosting efficiency \cite{cheung2022demonstration}. Fig. \ref{KAN_arch}(c) showcases nine randomly selected distinct nonlinear functions obtained through parameter tuning, demonstrating the versatility of the MRR nonlinearity. The RAMZI structure offers enhanced programmability over single MRRs by manipulating the phase of one MZI arm to achieve constructive and destructive interference at varying input powers. This enables efficient realization of diverse NAFs. Our experimentally demonstrated RAMZI design, fabricated at Advanced Micro Foundry (Fig. \ref{Measure}(a)), confirms the theoretical predictions \cite{peng2023all}. The device is fabricated on standard silicon on insulator (SOI) substrate. As shown in Fig. \ref{Measure}(b)-(d), three simulated (blue lines) and measured (black lines) responses are obtained with phase differences $\Delta \phi$. The tests were conducted with a gated optical signal to mitigate thermal nonlinearity, which can be further addressed by employing a low-loss platform. More data and figures can be found in \cite{peng2023all}. Notably, full programmability is achieved within a low input power range of 0-0.2 mW (-7 dBm), reducing laser power requirements and enhancing network efficiency. 

The ability to represent arbitrary nonlinear functions along edges is essential for the effective training of KANs \cite{liu2024kan}. The original digital implementation of the KAN uses highly parametric NAFs in the form of spline functions, often with more than a hundred trainable parameters per NAF. Consequently, representing arbitrary NAFs in analog KANs is a fundamental challenge for their effective implementation. In particular, the RAMZI unit is limited to just four parameters: two for adjusting MRR phase and loss, and two for controlling the phase shifter. While RAMZI units offer flexibility in creating functions, they have inherent limitations. To expand the range of available nonlinear functions and provide more control over network behavior, we propose a design where two RAMZI units are cascaded within a single unit cell called dual-RAMZI (D-RAMZI), depicted in Fig. \ref{KAN_arch}(b). An amplifier is used between the RAMZI units to guarantee sufficient power for triggering nonlinearity in the second RAMZI. The photonic KAN accelerator comprises D-RAMZI unit cells acting as edges between nodes. By directly connecting nodes through a single unit cell, this architecture minimizes the distance signals need to travel. This results in significant improvements in power efficiency and latency, enabling communication that is both fast and energy-efficient. A $N$×$M$ KAN consists of $NM$ D-RAMZIs with 9$\times$$NM+M$ parameters for tuning MRR phase, amplitude, MZI phase, and amplifiers. Correspondingly, the learnable parameters of one MLP layer and one MZI-ONN layer are $N^2$ and $2N^2$, respectively. Fortunately, KANs usually require much smaller network width and depth than MLPs.


\section{Results}
\label{Result}
In this section, we benchmark the ideal performance of the photonic KAN in simulation to characterize its expressivity given the limitations of the analog NAFs. In the next section, we will consider engineering challenges such as loss compensation and scalability. We employed PyTorch to implement the network architecture in our simulations, with architecture depicted in Figure \ref{KAN_arch}. The trainable D-RAMZI NAFs are parameterized using a semi-analytical approach. The MRR nonlinearity was precomputed using Equations \ref{eq4}-\ref{eq5}, with input power swept from 0 to 0.2 mW in 100 steps. To enable differentiation, this nonlinearity was interpolated by sweeping the loss and the phase in 16 steps, respectively. For the remaining D-RAMZI elements, we adopted an analytical approach based on S-matrices, with the phase shifters continuously tunable across a 0 to 2$\pi$ range.

\subsection{MNIST and Fashion-MNIST}

\begin{figure}[h!]
    \centering
    \includegraphics[scale=0.46]{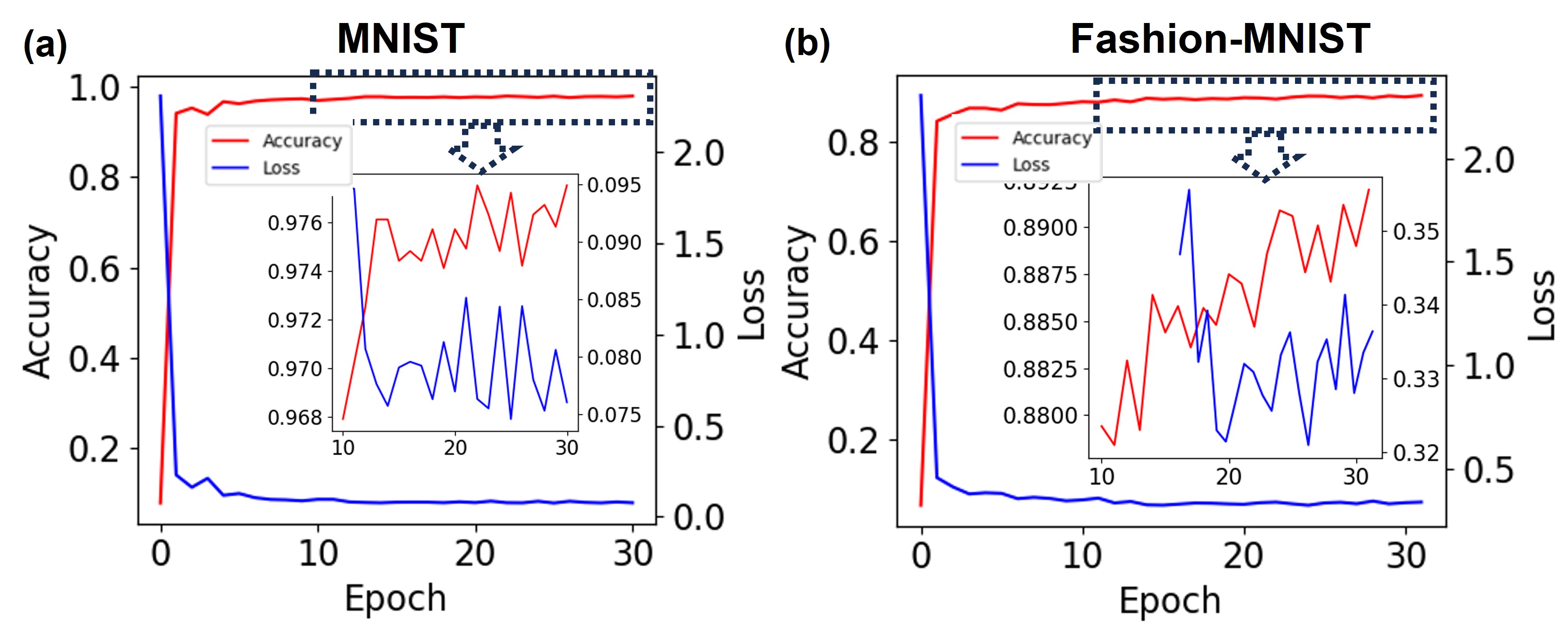}
    \caption{(a) MNIST results. (b) Fashion-MNIST results.}
    \label{MNIST}
\end{figure}

Our simulations utilized a two-layer photonic KAN network with [784, 64, 10] architecture, trained on the MNIST and Fashion-MNIST datasets. We employed Categorical Cross-Entropy loss, the AdamW optimizer with a learning rate of $1 \times 10^{-2}$, and an exponential learning rate scheduler (gamma 0.95) over 30 epochs. As shown in Fig. \ref{MNIST}, the network achieved a competitive 98\% accuracy on MNIST, and 89\% on the more demanding Fashion-MNIST dataset, which are both comparable to conventional KANs \cite{bodner2024convolutional, cheon2024demonstrating}. Our photonic architectures demonstrate rapid convergence, achieving over 80\% accuracy after a single training epoch. These results highlight the potential of photonic KANs in basic image classification tasks, with further improvement expected through additional training epochs and optimization.

\subsection{Photonic KANs are Prunable}

\begin{figure}[h!]
    \centering
    \includegraphics[scale=0.45]{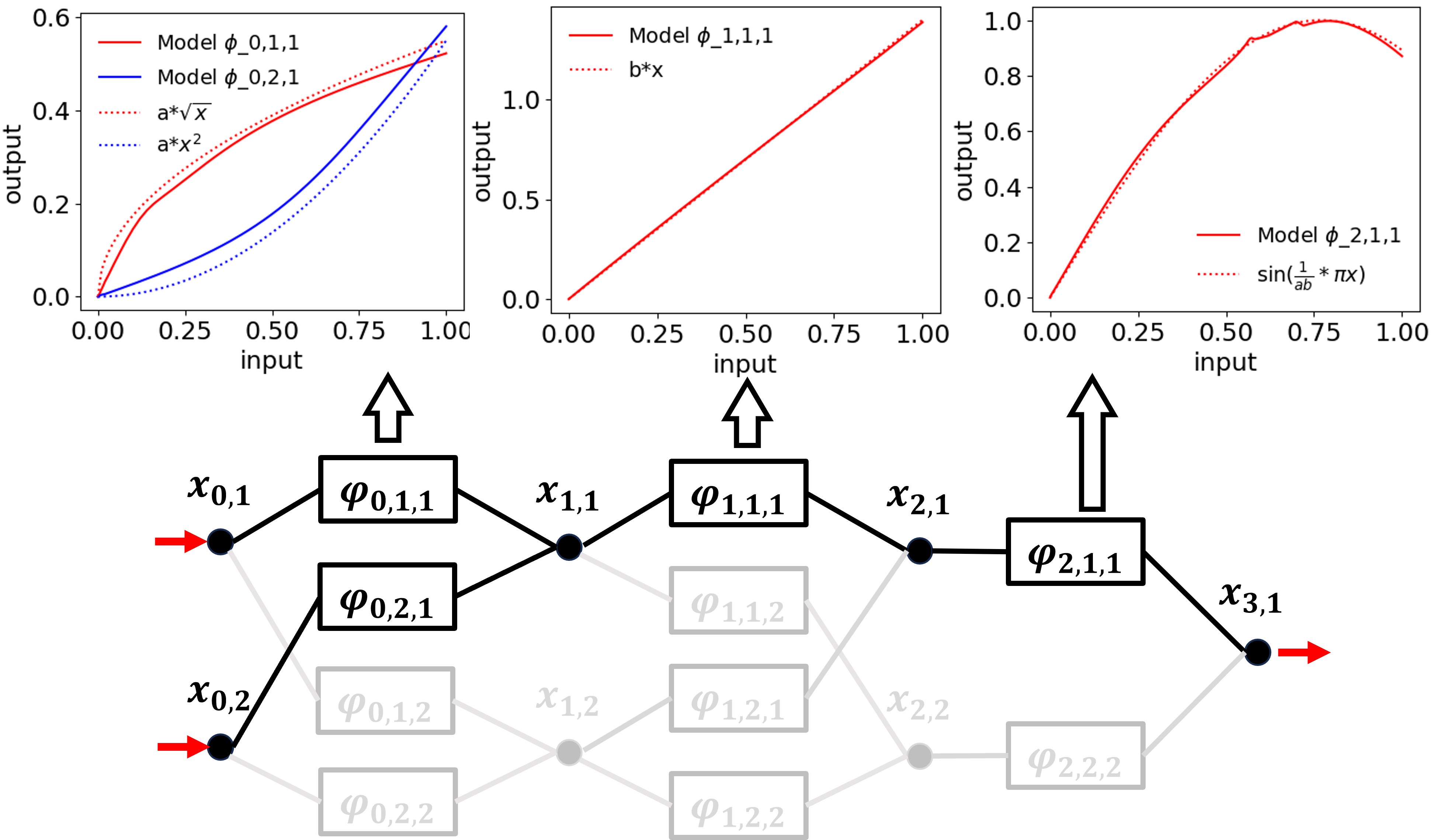}
    \caption{Pruned KAN interpret equations.}
    \label{Pruning}
\end{figure}

In practice, determining the network shape for a given dataset remains a challenge. We often lack a priori knowledge of the underlying function, making it difficult to predefine an ideal network structure. Ideally, approaches to determine this shape automatically would be desired, the idea is to start from a large enough overparameterized KAN and leverage sparsity regularization during training followed by pruning. This approach produces significantly more interpretable KANs compared to those without pruning. Furthermore, pruning significantly reduces model size and decreases hardware energy consumption.

Let us consider the simple function fitting task $f(x,y) = \sin\left(\pi\left(\sqrt{x} + y^2\right)\right)$, then we know that a [2, 1, 1] KAN is able to express this function. We train with sparsification regularization.

We used the same L1 regularization shown in \cite{liu2024kan}. The L1 norm of an activation function $\Phi$ of layer $l$ is defined to be its sum of learnable 1D functions matrix in Equation \ref{1}. In addition, the entropy of $\Phi_l$, $S(\Phi_l)$ is:
\begin{equation}
S(\Phi_l) = - \sum_{i=1}^{n_{\text{in}}} \sum_{j=1}^{n_{\text{out}}} \left| \frac{\phi_{i,j}}{|\Phi|_1} \right| \log \left( \left| \frac{\phi_{i,j}}{|\Phi|_1} \right| \right)
\label{2}
\end{equation}
We augment our regularization strategy by incorporating an additional coefficient, $A(\Phi_l)$, specifically targeting certain parameters, like the amplifier gain. This aims to drive these parameters towards zero, promoting sparsity and simplifying the model. Hardware-implemented KANs require a specialized regularization term to address the inherent variations across different hardware platforms. This term optimizes performance and stability by accounting for hardware-specific constraints. The total training objective $\ell_{\text{total}}$ is the prediction loss $\ell_{\text{pred}}$ plus L1, entropy regularization and coefficient regularization of all KAN layers. 
\begin{equation}
\ell_{\text{total}} = \ell_{\text{pred}} + \lambda \left( \mu_1 \sum_{l=0}^{L-1} L1(\Phi_l) + \mu_2 \sum_{l=0}^{L-1} S(\Phi_l)+ \mu_3 \sum_{l=0}^{L-1} A(\Phi_l) \right)
\label{3}
\end{equation}
where $\mu_1$, $\mu_2$ and $\mu_3$ are relative magnitudes, and $\lambda$ controls overall regularization magnitude.

We begin with a fully-connected [2, 2, 2, 1] KAN, uniformly sample 100 points in [0,1] and apply sparsification regularization during training to encourage the network to learn a sparse representation. Subsequent pruning, based on the score thresholds \cite{liu2024kan}, removes 'useless' nodes with weak incoming or outgoing connections. In this example, this process eliminates half of the neurons in the second and third layers. Visualizing the pruned network (Fig. \ref{Pruning}) reveals that functions with low magnitudes are effectively faded out, highlighting the important functional components. While our current method cannot remove entire layers due to stability issues, automatic pruning successfully simplifies the KAN to a [2, 1, 1, 1] structure. Importantly, the remaining activation functions visually resemble known symbolic functions ($\sqrt{x}$, $x^2$, $\sin{x}$) as shown in Fig. \ref{Pruning}, making it possible to correctly interpret the mathematical relationships captured by the model.  

In our photonic hardware implementation,  we first training on a digital GPU, followed by pruning of unnecessary connections. In the hardware itself, 'useless' nodes are deactivated by physically disconnecting them, which is setting their corresponding amplifier gains to zero for our specific architecture. We can easily achieve this by not pumping the amplifier, causing it to absorb light passing through it. This approach significantly reduces hardware energy consumption by eliminating power drain from inactive components. It is worth noting that the use of sparsification regularization, while improving model interpretability, does result in a slight decrease in accuracy. This is observed increase in MSE loss from $9.8 \times 10^{-5}$ for [2, 2, 2, 1] KAN without sparsification regularization to $1.2 \times 10^{-4}$ for [2, 2, 2, 1] KAN after pruning. This trade-off is understandable given that sparsification reduces the number of model parameters and the current unit cell design has limited tunability. However, we believe that if the unit cell could be designed to achieve arbitrary function representation, this accuracy issue could potentially be mitigated.

\subsection{Function Fitting}
We compare the performance of ideal MLPs, photonic KANs and conventional photonic MZI-based ONN, which typically relies on SVD techiques to represent MLP in the optical domain.  As shown in \cite{jacot2018neural}, MLPs perform poorly on high-frequency components, which are crucial for multi-scale partial differential equations (PDEs), image and audio compression, and medical applications. Neural Tangent Kernel (NTK) theory attributes this weakness to smaller eigenvalues in the NTK matrix. We compare the performance of our proposed photonic KANs against both ideal MLPs and conventional photonic MZI-based ONNs on function-fitting tasks involving high-frequency components. To ensure a fair comparison, we maintain consistent optimization methods and learning rates across all models.

\begin{figure}[h!]
    \centering
    \includegraphics[scale=0.5]{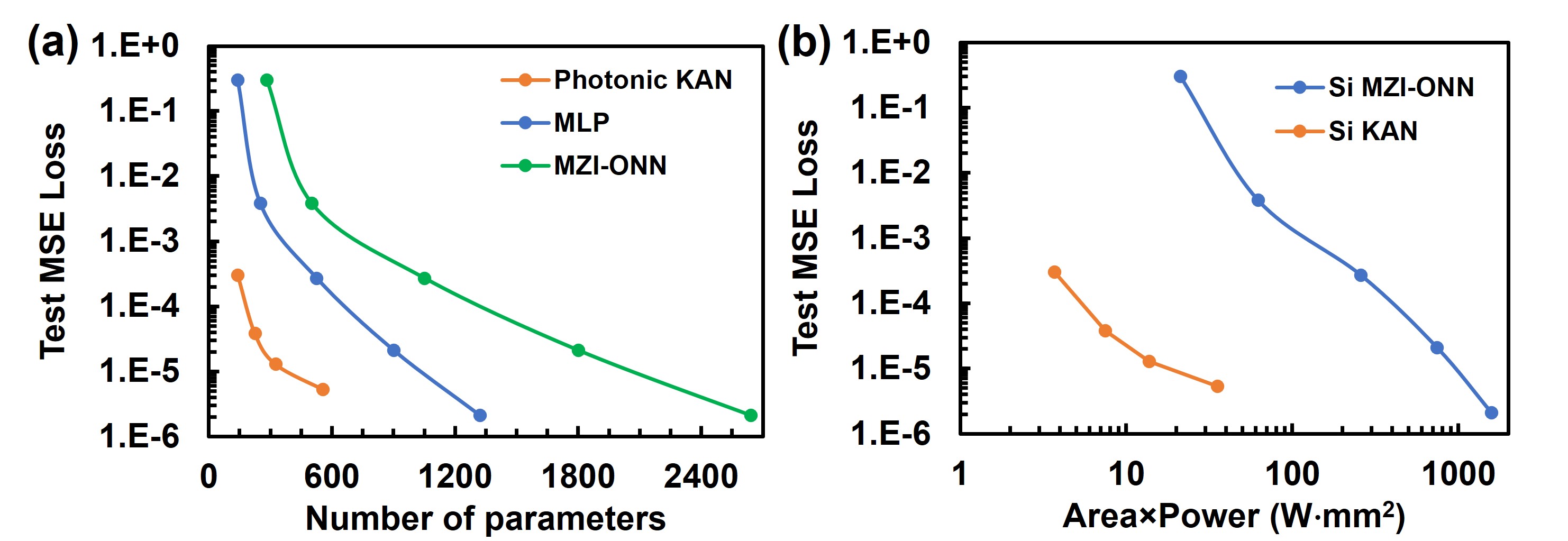}
    \caption{(a) Test loss for MLP and photonic KAN. (b) Footprint-energy efficiency for MLP and photonic KAN}
    \label{high_freq}
\end{figure}
We fitted function $y = \sin\left(\frac{\pi}{2} x\right) + 0.1 \sin(10 \pi x)$ using photonic KAN and MLP. As shown in  Fig. \ref{high_freq}, we increase KAN complexity from [1,3,3,1] to [1,5,5,5,1] and MLP complexity from [1,10,10,10,1] to [1,20,20,20,20,1], while maintaining the same dataset and optimizer (AdamW). As the results show, our photonic KAN demonstrates faster convergence and higher accuracy for problems with both low and high frequencies with the same parameters number. 

Due to the inherent unitary nature of MZI architectures, constructing MLPs directly within an MZI mesh is challenging. SVD techniques address this by decomposing weight matrices into two unitary and one diagonal matrix. However, this decomposition can lead to a doubling of the required tunable parameters. Ideally, an MZI-based ONN would perform similarly to an MLP layer given twice the number of tunable parameters. 


For this task, we also calculate the power consumption and area for photonic KAN and photonic MZI-based ONN using MZI Clement mesh. The photonic KAN improves the footprint-energy efficiency by around 65× achieving the similar accuracy due to the following reasons: 1. Reduced parameter requirements for the task. 2. Fewer MZIs, decreasing area and power consumption. 3. Shorter paths that significantly lower required laser power. The detailed calculations are provided in following section.

\section{Analysis of Photonic KAN Power, Efficiency, Footprint, and Latency}
We analyze the latency, power, area, and energy efficiency of our proposed photonic KAN. In static operation, power consumption primarily stems from six sources: laser wall-plug power, semiconductor optical amplifiers (SOAs) power, D-RAMZI mesh power, modulator circuit power, receiver circuit power, and analog-to-digital converters (ADCs)/digital-to-analog converters (DACs) power.

\begin{figure}[h!]
    \centering
    \includegraphics[scale=0.5]{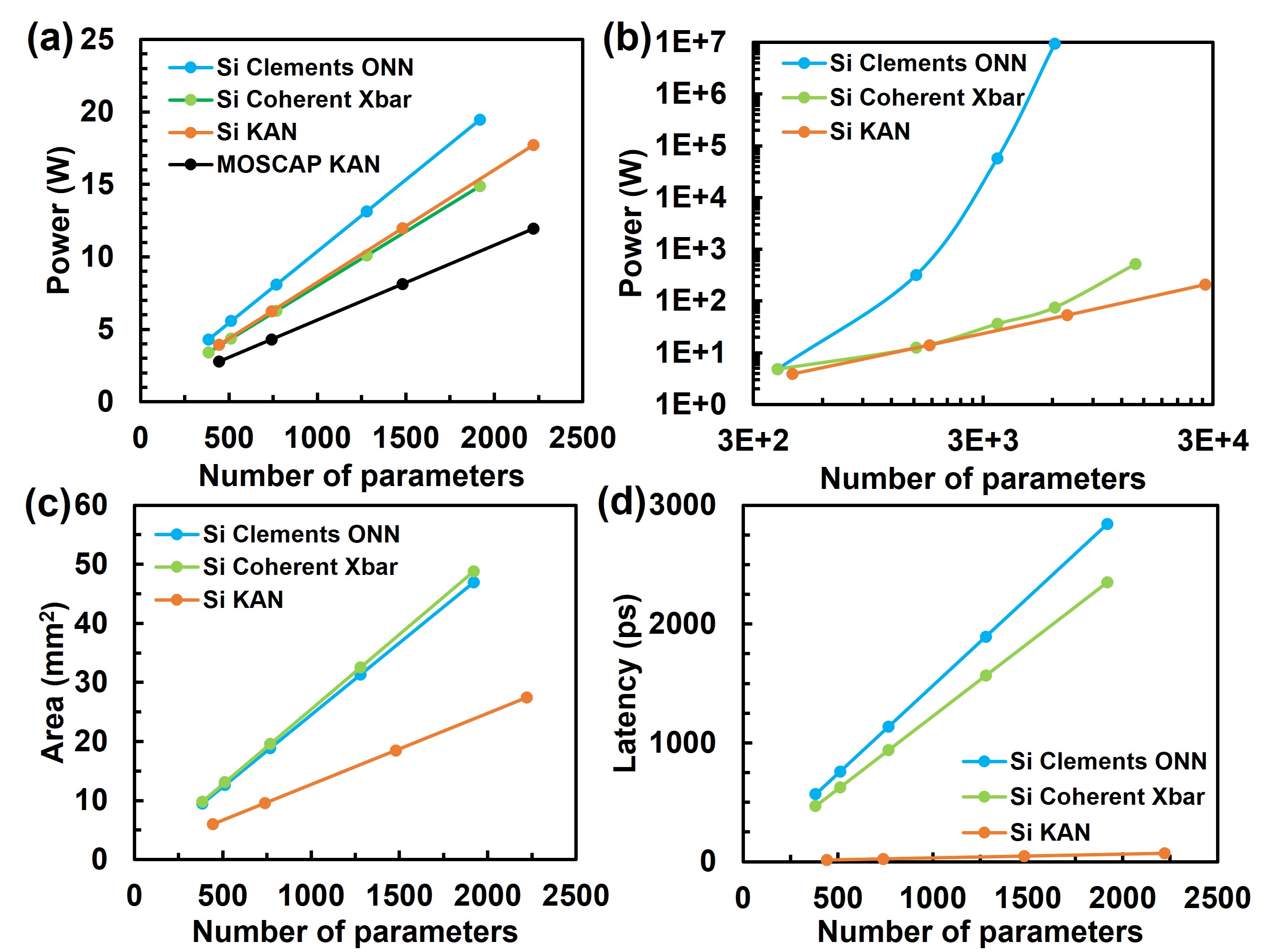}
    \caption{(a) Power consumption with parameter count (increasing depth). (b) Power consumption with parameter count (increasing width). (c) Footprint with parameter count (increasing depth). (d) Latency with parameter count (increasing depth).}
    \label{comp}
\end{figure}

\begin{figure}[ht!]
    \centering
    \includegraphics[scale=0.45]{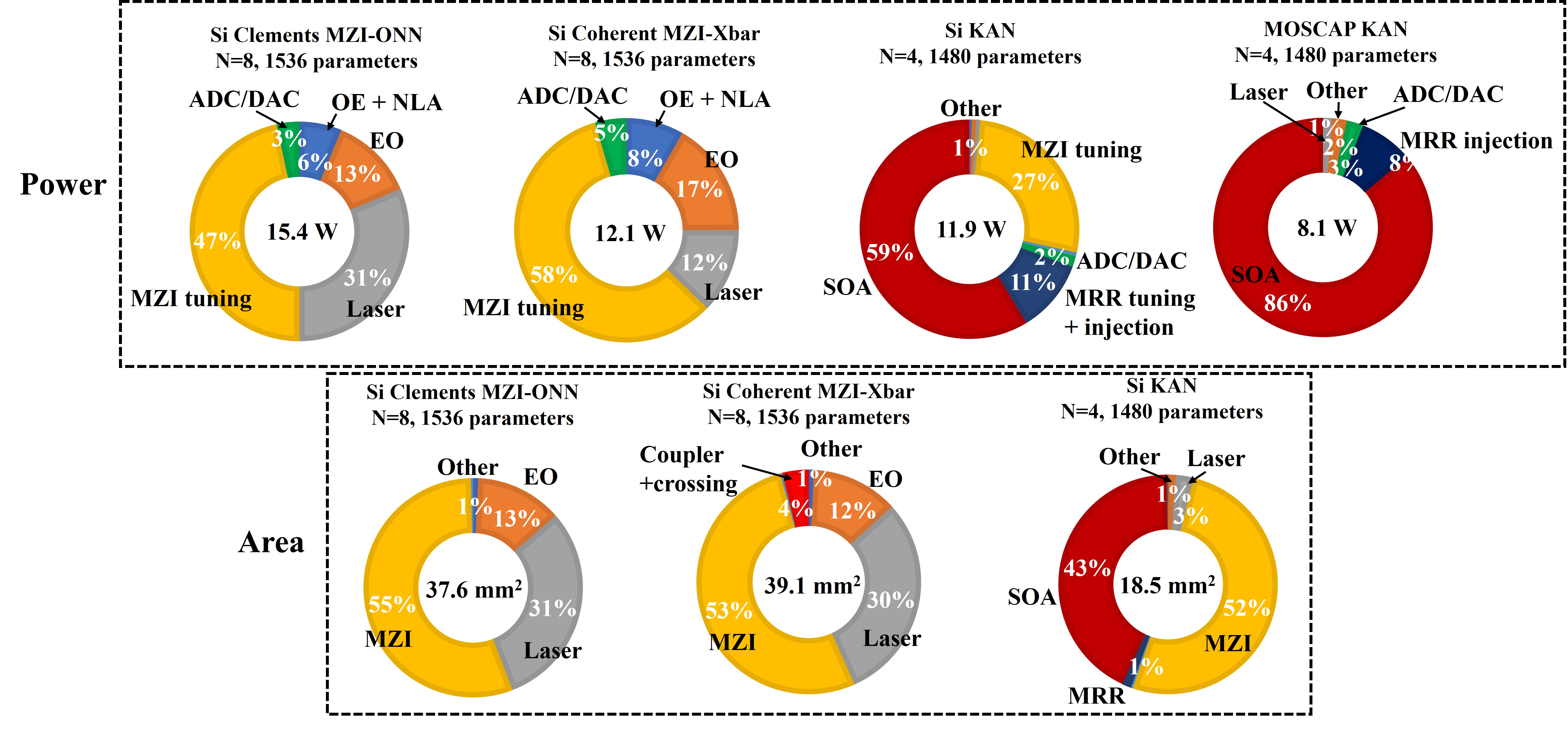}
    \caption{Power consumption and footprint for photonic ONN and KANs.}
    \label{pie}
\end{figure}
The presence of SOAs in the link to compensate for losses allows the laser to operate at the low power level needed to trigger the RAMZI nonlinearity. This threshold power is determined by our measurements, as detailed in Section 3. While on-chip SOAs are the primary power consumers, their operation in the low-gain and linear region minimizes power drive requirements. Additionally, the SOAs alleviate the laser's power burden, resulting in acceptable overall power consumption. RAMZI mesh power includes static contributions from MZI phase shifters, MRR tuning, and carrier injection. Transceiver power data comes from our demonstration and experiment at >25 GS operation, encompassing photonic device capacitance charging, drivers, transimpedance amplifiers (TIAs), power amplifiers, and clock power. Notably, our KAN architecture performs analog computation throughout without requiring OEO conversion. Our architecture only necessitates the use of high-speed DACs for modulating input signals and ADCs for detecting output signals. The power consumed by these DACs and ADCs, relative to the entire system, is lower than in conventional single-layer photonic accelerators. We determined passive device loss and footprint based on the process design kit (PDK) of a commercial foundry. The complete dataset used for these calculations is available in Table \ref{tab:Data}.

\begin{table}[h!]
    \centering
    \caption{Adopted component parameters for calculation.}
    \begin{tabular}{c c c c c c}
        \toprule
        Device & Power consumption (mW) & Footprint($mm^2$) & Insertion loss (dB) \\
        \midrule
        MZI phase shifter \cite{peng2023all}& 5 & 0.3$\times$0.05 & 1.2 \\
        MRR phase injection \cite{peng2023all}& 2 & 0.02$\times$0.02 & 0.2 \\
        Modulator+driver \cite{chang2023sub}& 20 & 1$\times$0.05 & 5.2 \\
        PD+driver \cite{chang2023sub,liu202356gb}& 10 & 0.003 & 0 \\
        Laser \cite{wang2019high}& (15\%wall-plug efficiency) & 0.3$\times$0.4 & 0 \\
        SOA \cite{shi2019deep}& 35 & 1$\times$0.05 & 0  \\
        ADC \cite{liu202210gs} & 15 & 0.003 & 0 \\
        DAC \cite{caragiulo2020compact} & 50 & 0.011 & 0 \\
        OEO conversion+ Nonlinear function \cite{williamson2019reprogrammable}& 30 & 0.053 & 5.2 \\
        1:2 Splitters/2:1 Combiners \cite{tan2023foundry}& 0 & 0.05$\times$0.01 & 0.1 \\
        Crossing \cite{tan2023foundry}& 0 & 0.008$\times$0.008 & 0.1 \\
        Waveguide \cite{tan2023foundry}& 0 & / & 0.3\,\text{dB/mm}  \\  
        \bottomrule
    \end{tabular}
    \label{tab:Data}
\end{table}

We compare our KAN with the conventional Clements's MZI ONN \cite{pai_matrix_2019} and advanced coherent MZI-Xbar ONN \cite{giamougiannis2023coherent}. Figure \ref{comp} shows that photonic KANs have comparable power consumption to the coherent MZI-Xbar, and both demonstrate a clear advantage over photonic MZI-ONNs in terms of power efficiency when using the same number of parameters. Fig.\ref{pie} further details the power consumption breakdownfor each architecture. When N=8, both MZI-ONNs power usage are primarily dominated by MZI tuning, laser operation, and OE conversion. In contrast, SOAs constitute the major power consumer in photonic KANs. Notably, encoding and decoding costs (DAC and ADC) are minimal due to the deep network structure. As the network depth increases (with a fixed small width), the photonic KAN consistently consumes less power than conventional Clements's MZI ONN. This power saving of approximately 25\% is achieved through two primary factors: the utilization of fewer MZI devices and reduced laser power coming from less optical path loss. Furthermore, by adopting a MOSCAP platform, where the static power consumption of MZI tuning is zero, we can potentially reduce the power consumption of the photonic KAN to half that of a conventional MZI-based ONN. 

The architectural differences between photonic KANs, coherent MZI-Xbar ONNs and conventional photonic MZI ONNs lead to distinct scaling properties. Conventional $N$×$N$ MZI ONN meshes require $N^2$ MZIs with a maximum optical path of (2$N$ + 1) MZIs. This leads to exponential increases in path loss and laser power with growing network width. Conversely, photonic KANs transmit through only one D-RAMZI unit per connection, and coherent MZI-Xbar ONNs transmit through one MZI, ensuring a fixed optical path loss irrespective of network size. While larger networks do require more combiners/splitters, the resulting loss is minor and scales logarithmically with network width, easily compensated by amplifier gain. Consequently, as Fig.\ref{comp}(b) demonstrates, photonic KAN power consumption scales linearly with network size, while conventional MZI ONN power consumption exhibits exponential growth. This linear scaling makes photonic KANs a promising solution for larger neural networks. However, coherent MZI-Xbar ONNs incur additional losses from couplers and crossings that increase with network width. This explains the lower power consumption observed in our KAN compared to coherent MZI-Xbar ONNs. The significant number of crossings in MZI-Xbar ONNs can lead to severe crosstalk issues. Furthermore, photonic KANs usually require less network depth compared to MZI-based ONNs to achieve the similar accuracy. 

In terms of footprint, photonic KANs offer a distinct advantage over MZI-based ONNs. This is due to the higher parameter density of RAMZI units. Each RAMZI unit, while occupying a similar footprint to a conventional MZI unit, contains twice the number of tunable parameters (four versus two). Consequently, as illustrated in Fig.\ref{comp}, a photonic KAN can achieve the same overall number of parameters as an MZI mesh while occupying roughly half the physical space. Additionally, coherent MZI-Xbar ONNs have an even larger footprint than conventional MZI ONNs due to the substantial number of additional combiners and crossings required. As demonstrated before, our KANs are amenable to pruning. This allows us to further reduce the photonic KAN's footprint and power consumption while preserving good accuracy.

Reducing latency in ONNs is crucial for unlocking their full potential in real-world applications. The optics latency increases approximately linearly with the size as the optical path increases. The EO/OE latency remains almost the same for each layer, which accounting for the contributions from the group delay of the optical-to-electrical conversion stage, the delay associated with the nonlinear signal conditioner and the RC time constant of the phase modulator. Our photonic KAN design achieves a remarkable 50× reduction in overall latency by eliminating the need for multiple EO/OE conversions and enabling short optical path. Compared to the proposed all-optical ONN \cite{peng2023all}, our KAN also exhibits significantly lower latency. The all-optical ONN suffers from substantial optical latency for large-size network and necessitates much deeper network requirements.


\section{Conclusion}
We have proposed the novel photonic neural network architecture inspired by the KAN. We showcase its performance in function fitting and image classification, highlighting its potential for broader applications. This design overcomes some fundamental constraints of existing photonic neural network in following aspects:

\begin{enumerate}
    \item \textbf{Improved accuracy and convergence:} Photonic KANs exhibit superior performance compared to photonic MLPs in terms of both accuracy and convergence speed across a range of frequencies. Notably, Photonic KANs maintain this performance with half the parameter count, resulting in smaller model sizes and reduced computational requirements.
    \item \textbf{Efficiency:} Photonic KAN achieves around 2× energy, 2× area and 50× latency reductions compared to prior conventional MZI-based photonic accelerators as simulation with the same parameters.
    \item \textbf{Scalability: } A key advantage of photonic KANs is their linear power scaling with network width, overcoming the exponential growth limitations of other photonic accelerators and paving the way for significantly larger and more powerful photonic neural networks.
    \item \textbf{Interpretability:} Photonic KAN show enhanced model understanding which MLP do not have. This is crucial for building trust in critical applications and ensuring fair and ethical use, as it allows for easier identification and correction of errors, biases, or unexpected behaviours.
\end{enumerate}

Our initial findings highlight the promise of photonic KANs, outperforming MLPs in scientific tasks like function fitting and image classification. However, further development is needed to overcome two key challenges:

\begin{enumerate}
    \item \textbf{Training Speed Bottleneck:} Our photonic KANs training speed is 6x slower than MLPs on GPUs, given the same number of parameters. This limitation arises from the inability to leverage batch computation for the diverse range of activation functions used in KANs. The potential for on-chip training presents an interesting avenue for future exploration. It's uncertain whether this approach will encounter bottlenecks similar to those seen in ONNs.
    \item \textbf{Limited Functional Flexibility:} Photonic devices, due to their physical constraints, currently struggle to achieve the same level of flexibility in representing arbitrary nonlinear functions as digital electronic circuits. This limitation hampers the full potential of photonic KANs and necessitates further research into more versatile photonic components. Wavelength division multiplexing (WDM) may offer a solution by providing an additional degree of freedom. On-chip quantum dot (QD) comb lasers offer multiple wavelengths, each imprinted with input vectors via modulators. Subsequently, multi-wavelength D-RAMZIs with multiple MRRs provide distinct nonlinear responses per wavelength, expanding the range of available functions.
\end{enumerate}

While photonic KANs have not yet surpassed electronic accelerators in large-scale demonstrations, overcoming the current challenges in training speed and functional flexibility will unlock their full potential for a wide range of scientific and practical applications. Our proposed novel architecture offers a new pathway for increasing the extensibility of photonic neural networks. While this is the first proposed structure, other on-chip or free-space methods utilizing tunable nonlinear photonics could also be used to implement KAN. We envision photonic KANs as a future cornerstone of AI for solving PDEs and other machine learning tasks, where they may eventually replace the current standard of photonic neural networks \cite{yu2024kan}.


\bibliographystyle{unsrt}  
\bibliography{references}

\end{document}